\documentclass{article}
\usepackage[english]{babel}
\usepackage[table,dvipsnames,svgnames]{xcolor}
\usepackage[letterpaper, top=2cm, bottom=2cm, left=2.6cm, right=2.6cm, marginparwidth=3.0cm, marginparsep=0.5cm]{geometry}
\usepackage{setspace}
\usepackage{amsmath}
\usepackage{amsthm}
\usepackage{bm}
\usepackage{float}
\usepackage{booktabs}
\usepackage{graphicx}
\usepackage{dcolumn}
\usepackage{tabularx}
\usepackage{threeparttable}
\usepackage{longtable}
\usepackage{array}
\usepackage{caption}
\usepackage{ragged2e}
\usepackage{multirow}
\usepackage{tikz}
\usepackage{placeins}
\usepackage{enumitem}
\usepackage{authblk}
\usepackage[normalem]{ulem}

\definecolor{sugg}{HTML}{1CAF2E}

\definecolor{bar_random}{HTML}{579765}
\definecolor{bar_consist}{HTML}{DD6E92}
\definecolor{bar_optimal}{HTML}{0090F4}
\newcolumntype{L}[1]{>{\raggedright\let\newline\\arraybackslash\hspace{0pt}}m{#1}}
\newcolumntype{C}[1]{>{\centering\let\newline\\arraybackslash\hspace{0pt}}m{#1}}
\usepackage[numbers,square,sort&compress]{natbib}
\setcitestyle{numbers,square,comma}
\usepackage[colorlinks=true, allcolors=blue]{hyperref}

\begin{document}
\title{\textbf{Rethinking Sampling Strategy in Link Prediction}}
\author{Yilin Bi, Zhenyu Deng, Xinshan Jiao, Tao Zhou$^{*}$}
\affil{CompleX Lab, School of Computer Science and Engineering, University of Electronic Science and Technology of China, Chengdu 610054, China; \\ $^{*}$Corresponding author: zhutou@ustc.edu}
\maketitle

\begin{abstract}  
Many real-world networks are incomplete, making link prediction a fundamental challenge in network science. To train parameters and evaluate algorithms, observed links are usually divided into three subsets, namely training, validation, and probe sets. This division implicitly involves two sampling processes: first-stage sampling yields the probe set and second-stage sampling obtains the variation set. To date, our understanding of how these two sampling processes affect algorithm performance remains quite limited. To address this issue, we propose a sampling scheme called $\beta$-sampling, where the sampling probability of a link is proportional to the product of the degrees of its two endpoints raised to the power of $\beta$. Experiments on 45 real-world networks reveal that the structural characteristics of missing links, as simulated via varying probe sets, substantially impact prediction accuracy. When missing links tend to connect high-degree nodes, such links can be predicted accurately with ease. Furthermore, even with a fixed probe set, second-stage sampling still exerts a significant influence on prediction accuracy. Notably, the optimal second-stage sampling strategy differs from \textit{random sampling} (which randomly selects links to form the validation set) and \textit{consistent sampling} (which guarantees that links in the validation and probe sets share identical structural characteristics). 
\end{abstract}

\section{Introduction}
Real-world systems are rich in interactions. Social relationships \cite{vega2007} and biomolecular interactions \cite{gosak2018} are typical examples. These systems can be represented as networks composed of nodes and links. Such representation provides a unified framework for describing how entities are linked, offering a powerful tool to analyze the structure, function and evolution of complex systems \cite{strogatz2001, mitchell2009, barabasi2016, newman2018}. In practice, many real-world networks suffer from incompleteness \cite{kossinets2006}. Recovering missing links from imperfect network data thus constitutes a core challenge in network science, widely known as the link prediction problem \cite{lv2011, wang2015, martinez2016, kumar2020, zhou2021, wu2022, xu2026}. Link prediction has found successful applications in evaluating network evolution \cite{wang2012,zhang2015}, recommender systems \cite{lv2012, bobadilla2013}, biological data mining \cite{bi2022, musawi2025}, and so on.

Over the past two decades, link prediction has developed rapidly. Early studies mainly relied on similarity-based heuristics, such as Common Neighbors index \cite{liben2007}, Adamic-Adar index \cite{adamic2003}, and Resource Allocation index \cite{ou2007, zhou2009}. Later, maximum likelihood methods based on global structure were introduced, including the Hierarchical Structure Model \cite{clauset2008} and the Stochastic Block Model \cite{guimera2009}. Recently, graph representation learning and graph neural networks \cite{scarselli2008} have further pushed link prediction toward end-to-end learning, with representative examples including GraphSAGE \cite{hamilton2017}, Graph Attention Networks \cite{velivckovic2018}, and Variational Graph Autoencoders \cite{kipf2016}. Yet despite the diversity of prediction algorithms, their evaluation still relies on a largely unified protocol \cite{jiao2025}: the observed links are split into three disjoint subsets: \textit{training set} $E^T$, \textit{validation set} $E^V$, and \textit{probe set} (also called test set) $E^P$, where the training set is used to calculate structural features, the validation set supports parameter tuning, and the probe set is held out for evaluation. 

The above partition actually involves two distinct sampling processes. In the first stage, the original link set is split into a retained observed set $E^T \cup E^V$ and a probe set $E^P$, with the later mimicking the missing links to be recovered. Studying this stage allows us to examine whether different structural characteristics of missing links (named as \textit{missing patterns}) affect prediction accuracy and how large this effect can be. In real-world networks, missing links are usually not randomly distributed \cite{he2024}. Zhu \textit{et al.} \cite{zhu2012} found that real missing links are more likely to occur between low-degree nodes, and algorithms performing well for randomly distributed missing links may fail subject to real missing patterns. Ahmed \textit{et al.} \cite{ahmed2012} reported that different first-stage sampling strategies will lead to significantly different performance rankings of prediction algorithms. These findings suggest that missing links may exhibit domain-specific and non-random structural patterns, rather than being well represented by a randomly sampled subset of observed links, and such missing pattern will considerably affect prediction accuracy.

In the second stage, the retained observed links are further divided into $E^T$ and $E^V$. Studying this stage allows us to check whether the training--validation split affects algorithmic performance given the probe set. Neville \textit{et al.} \cite{neville2009} pointed out that relational data naturally violate the independent and identically distributed assumption, so standard cross-validation may fail on graph data. He \textit{et al.} \cite{he2024} compared multiple sampling strategies, showing that algorithmic performance depends jointly on the sampling strategy and network characteristics, and that no universally optimal sampling strategy exists. Notice that, these studies do not explicitly separate the sampling processes in two stages. In particular, they leave open whether the second-stage sampling still affects prediction performance once the probe set has been specified. Furthermore, a natural design goal of link sampling strategies is to align the structural characteristics of links within the validation set with those of missing links, so as to mitigate performance degradation when algorithms generalize from the training objective (validation set) to the prediction objective (probe set). This paper will verify whether this conjecture holds true.

To address these issues, we propose a sampling strategy with a tunable parameter $\beta$, named as $\beta$-sampling. Under $\beta$-sampling, the probability that a link with two endpoints $v_i$ and $v_j$ is sampled is proportional to $(k_i \cdot k_j)^\beta$, where $k_i$ and $k_j$ are the degrees of $v_i$ and $v_j$. This setting allows us to study the above-mentioned problems under controlled conditions. More specifically, if we denote the parameters in the first-stage and second-stage sampling as $\beta_1$ and $\beta_2$, then by varying $\beta_1$ we can examine whether the change of the probe set affects algorithmic performance, and when $\beta_1$ is fixed, we can compare the performance corresponding to random sampling ($\beta_2=0$) and consistent sampling ($\beta_2=\beta_1$) in the second stage. Through extensive experiments on $45$ real-world networks, we show that the parameter $\beta_1$ largely affects the algorithmic performance, indicating that the missing patterns matter. We further demonstrate that, with a fixed probe set, random sampling and consistent sampling achieve competitive prediction accuracy, and neither of them is the best-performing strategy. These findings suggest that link sampling is not an auxiliary experimental detail in link prediction, but a key factor that affects prediction difficulty and attainable performance.

\section{Methods}
Consider a simple undirected network $G=(V,E)$ with no self-loops, where $V=\{v_1, v_2, \dots, v_N\}$ denotes the node set and $E$ denotes the link set. Let $\bm{A}\in\{0,1\}^{N \times N}$ represents the adjacency matrix of $G$, where $a_{ij} = 1$ if node $v_i$ and $v_j$ are linked, and $a_{ij} = 0$ otherwise. The universal link set is defined as $U=\{(v_i,v_j) \mid v_i,v_j\in V,\ i<j\}$, which contains all possible node pairs. The observed link set $E$ is partitioned into three mutually disjoint subsets, as $E = E^T \cup E^V \cup E^P$ and $E^T \cap E^V = E^T \cap E^P = E^V \cap E^P = \emptyset$. For parameterized algorithms, $E^T$ is used to construct the training network, $E^V$ is used for parameter training, and $E^{P}$ is kept hidden and used only for performance evaluation \cite{jiao2025}. The algorithm is required to predict links from $U \setminus (E^{T} \cup E^{V})$, among which links in $E^P$ are treated as positive samples. By contrast, for parameter-free algorithms, such as Common Neighbors index and Resource Allocation index, no validation set is required for parameter tuning. Therefore, the union $E^T \cup E^V$ is used for score calculation, and the probe set $E^P$ is used for evaluation. The standard evaluation procedure is illustrated in \autoref{fig_process}(a).

\subsection{$\beta$-sampling}
\begin{figure}[htbp]
	\centering
    \centerline{\includegraphics[width=1.00 \linewidth]{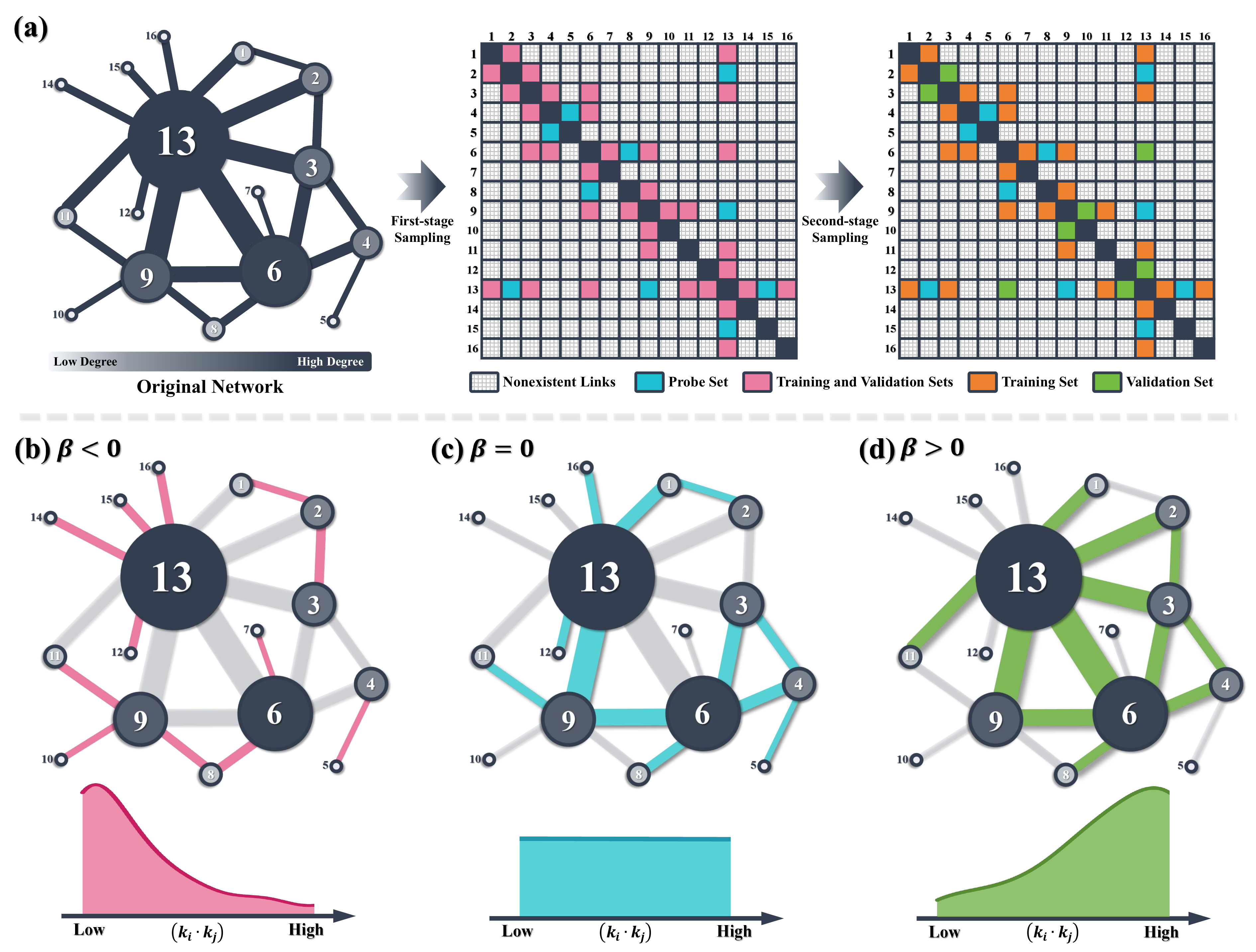}}
	\caption{{\bf Illustration of the two-stage $\beta$-sampling processes.} (a) The workflow of the two sampling processes. In the first-stage sampling, links in $E^T \cup E^V$ are sampled out from $E$ and the remaining links constitute the probe set $E^P$. In the second-stage sampling, links in $E^T$ are sampled out from $E^T \cup E^V$ and the remaining links belong to the validation set $E^V$. (b)-(d) demonstrate the effect of $\beta$ on link selection. With a fixed sample size of $12$ links out of $22$ total links, each plot presents a distinct scenario governed by $\beta$. The top panel visualizes the sampled links (colored edges), while the bottom panel displays the distribution of node degree products for these sampled links. Specifically, (b) represents $\beta < 0$, favoring links between nodes with lower degree products; (c) corresponds to $\beta =0$, where each link has an equal probability to be sampled; and (d) depicts $\beta > 0$, preferring links between node with higher degree products.}
	\label{fig_process}
\end{figure}

For any link $(v_i,v_j)$, its sampling probability under $\beta$-sampling is defined as
\begin{equation}
	p_{ij}(\beta) \propto (k_i k_j)^{\beta},
\end{equation}
where $\beta$ is a tunable parameter controlling the structural bias of sampling. When $\beta > 0$, links between nodes with higher degree products are more likely to be selected, and when $\beta < 0$, sampling is biased toward links between nodes with lower degree products. The case $\beta=0$ recovers the standard random sampling (see \autoref{fig_process}(b)--(d)).

We apply $\beta$-sampling twice to partition the link set $E$. In the first stage, we sample $f|E|$ links from $E$ according to $p_{ij}(\beta_1)$ to form $E^T \cup E^V$, and the unselected links constitute the probe set $E^P$. In the second stage, we sample  $f \cdot |E^T\cup E^V|$ links from $E^T\cup E^V$ according to $p_{ij}(\beta_2)$ to construct the training set $E^T$. The remaining links form the validation set $E^V$. Here, $f$ is the sampling fraction, fixed as $0.8$ in the later experiments.

\subsection{Algorithms}
To examine the impact of $\beta$-sampling on link prediction performance, we implemented eight representative link prediction algorithms, including four parameter-free algorithms and four parameterized algorithms. These algorithms cover multiple levels of structural information, ranging from local and quasi-local patterns to global topology. For each candidate $(v_i,v_j)$, each algorithm assigns a score $s_{ij}$, which reflects the propensity to be a missing link. A higher $s_{ij}$ is therefore interpreted as evidence that the candidate is a true missing link. The scoring functions and original references for these algorithms are detailed in \autoref{tab:algorithms}.

\begin{table}[htbp]
\centering
\caption{\textbf{Summary of link prediction algorithms.} Here, $s_{ij}$ is the prediction score for node pair $(v_i, v_j)$. $\mathbf{A}$, $\mathbf{D}$, $\mathbf{L}=\mathbf{D}-\mathbf{A}$, and $\mathbf{I}$ denote the adjacency matrix, degree matrix, Laplacian matrix, and identity matrix, respectively. The degree matrix is a diagonal matrix whose diagonal entries correspond to the degrees of each node in a graph, with all off-diagonal elements equal to zero. $\Gamma(i)$ is the neighbor set of node $v_i$. In CH2, $d_z^{\mathrm{int}}$ and $d_z^{\mathrm{ext}}$ are the internal and external degrees of $z$. In LRW, $\pi_{ij}(t)$ is the $t$-step transition probability from $v_i$ to $v_j$ (we set $t=3$ in the later experiments). In TSAA, $\mathbf{S}^{AA}$ denotes the Adamic--Adar similarity matrix with entries $s_{ij}^{AA}$. The algorithm-specific parameters are $\alpha_{\mathrm{MFI}}>0$, $\alpha_{\mathrm{Katz}}\in(0,1/\lambda_1)$, where $\lambda_1$ is the spectral radius of $\mathbf{A}$, and attenuation parameters $\alpha_{\mathrm{LP}}$ and $\epsilon$.}
\label{tab:algorithms}
\renewcommand{\arraystretch}{1.6} 
\resizebox{\textwidth}{!}{
\begin{tabular}{llc}
\toprule
\textbf{Algorithm} & \textbf{Scoring Function} & \textbf{Reference} \\
\midrule

Common Neighbors (CN) & 
$s_{ij}^{\text{CN}} = |\Gamma(i) \cap \Gamma(j)|$. & \cite{liben2007} \\

Resource Allocation (RA) & 
$s_{ij}^{\text{RA}} = \sum_{z \in \Gamma(i) \cap \Gamma(j)} \frac{1}{k_z}$. & 
\cite{ou2007, zhou2009} \\

Matrix Forest Index (MFI) & 
$s_{ij}^{\text{MFI}} = [(\mathbf{I} + \alpha_{\text{MFI}} \mathbf{L})^{-1}]_{ij}$. & 
\cite{chebotarev2006} \\

\begin{tabular}[c]{@{}l@{}}L2-based Cannistraci-Hebb network \\ automaton model number two (CH2)\end{tabular} & 
$s_{ij}^{\text{CH2}} = \sum_{z \in \Gamma(i) \cap \Gamma(j)} \frac{1 + di_z}{1 + de_z}$. & \cite{muscoloni2023} \\

Local Path Index (LP) & 
$s_{ij}^{\text{LP}} = (\mathbf{A}^2 + \alpha_{\text{LP}} \mathbf{A}^3)_{ij}$. & 
\cite{zhou2009, lv2009} \\

Local Random Walk (LRW) & 
$s_{ij}^{\text{LRW}} = k_i \cdot \pi_{ij}(t) + k_j \cdot \pi_{ji}(t)$. & \cite{liu2010} \\

Katz & 
$s_{ij}^{\text{Katz}} = [(\mathbf{I} - \alpha_{\text{Katz}} \cdot \mathbf{A})^{-1} - \mathbf{I}]_{ij}$. & \cite{katz1953} \\

Transferring Similarity Adamic-Adar (TSAA) & 
\begin{tabular}[c]{@{}l@{}}$s_{ij}^{\text{TSAA}} = [(\mathbf{I} - \epsilon \mathbf{S}^{\text{AA}})^{-1} \mathbf{S}^{\text{AA}}]_{ij}$, \\ $s_{ij}^{\text{AA}} = \sum_{z \in \Gamma(i) \cap \Gamma(j)} \frac{1}{\log(k_z)}$. \end{tabular} & \cite{sun2009} \\

\bottomrule
\end{tabular}}
\end{table}

\subsection{Evaluation Metrics}
For a given algorithm, we calculate a score $s_{ij}$ for each candidate $(v_i,v_j)$ in $U \setminus (E^{T} \cup E^{V})$ and rank all candidates in descending order. Following previous studies on evaluation metrics in link prediction \cite{zhou2023, jiao2024, bi2024, wan2025}, we select three representative metrics to evaluate algorithmic performance in this study.

\textbf{Area Under the ROC curve (AUC)} quantifies the area under the Receiver Operating Characteristic (ROC) curve \cite{hanley1982}, serving as a standard metric to evaluate the model's ability to distinguish between positive and negative samples. Let $r_i$ denote the rank of the $i$-th true missing link in the predicted list of $U \setminus (E^{T} \cup E^{V})$ candidates, and $\langle r\rangle=\sum_{i=1}^{|E^{P}|}\frac{r_{i}}{|E^{P}|}$ be the average rank, the AUC value is:
 \begin{equation}
	AUC=\frac{1}{|E^{P}|}\sum\limits_{i=1}^{|E^{P}|}(1-\frac{r_i-i}{|U \setminus E|})=1-\frac{\langle r\rangle}{|U \setminus E|}+\frac{|E^{P}|+1}{2 \cdot |U \setminus E|}.
 \end{equation}
This value can be interpreted as the probability that a randomly selected true missing link in $E^P$ is assigned a higher score than a randomly selected nonexistent link in $U \setminus E$. Considering that in link prediction, the number of nonexistent links far exceeds the number of missing links in the probe set, we adopt the widely used approximation \cite{zhou2023}:
\begin{equation}
	 AUC \approx 1-\frac{\langle r\rangle}{|U \setminus (E^{T} \cup E^{V})|}.
\end{equation}

\textbf{Area Under the Precision-Recall curve (AUPR)} measures the area under the Precision-Recall (PR) curve \cite{davis2006}, which is particularly suitable for scenarios with extreme class imbalance. The AUPR is computed as: 
\begin{equation}
	 AUPR=\frac{1}{2 \cdot |E^{P}|}(\sum\limits_{i=1}^{|E^{P}|}\frac{i}{r_i}+\sum\limits_{i=1}^{|E^{P}|}\frac{i}{r_{i+1}-1}).
\end{equation}

\textbf{Normalized Discounted Cumulative Gain (NDCG)}, a metric originating from information retrieval, evaluates ranking quality by assigning higher scores to latent links that appear earlier in the ranking list \cite{Jarvelin2002}. It is defined as the ratio of Discounted Cumulative Gain (DCG) to the Ideal DCG (IDCG):
\begin{equation}
	 NDCG=\frac{DCG}{IDCG}=\frac{\sum_{i=1}^{|E^{P}|}\frac{1}{\log_2(1+r_i)}}{\sum_{r=1}^{|E^{P}|}\frac{1}{\log_2(1+r)}}.
\end{equation} 
 
In our experiments, AUC, AUPR, and NDCG exhibite consistenttrends across datasets; therefore, to save space, we report results measured by AUC in the main text, while results for AUPR and NDCG are provided in the \textbf{Appendix}.

\subsection{Data}
We employ $45$ real-world networks spanning social, biological, technological, and informational domains to ensure broad representativeness \cite{bi2025}. These networks vary substantially in scale and connectivity, with the number of nodes $N$ ranging from $21$ to $1,882$ and the number of links $M$ from $39$ to $7,513$. Their structural properties also differ widely. The average shortest path length $\langle L \rangle$ spans from $1.32$ to $18.40$, indicating that some networks have compact global structures, such as \textit{Bison} and \textit{Japanese-macaques}, whereas others exhibit much longer characteristic path lengths, such as \textit{inf-euroroad}. The Clustering Coefficient (CC) \cite{watts1998} varies from $0.00$ to $0.79$, reflecting substantial differences in local clustering. The assortativity coefficient $r$ \cite{newman2002} ranges from $-0.65$ and $0.46$, covering both strongly disassortative networks, such as \textit{Kenyan-Households-Across}, and clearly assortative networks, such as \textit{netscience}. Overall, these $45$ networks provide a diverse empirical basis for evaluating link prediction algorithms, with detailed statistics presented in \autoref{tab:data}.

\captionsetup{font=small} 
\begin{table}[htbp]   
\centering
\caption{\textbf{Topological properties for the $45$ real-world networks used in this study}. $N$ denotes the number of nodes, $M$ is the number of links, $\langle k \rangle$ is the average degree, $\langle L \rangle$ represents the average shortest path length, $CC$ refers to the clustering coefficient, $d$ is the diameter, $r$ denotes the assortativity.}
\label{tab:data}
\footnotesize  
\setlength{\tabcolsep}{4pt} 
\renewcommand{\arraystretch}{1.1} 
\begin{tabular}{l c c c c c c c}
\toprule
\textbf{Networks} & 
\textbf{$N$} & 
\textbf{$M$} & 
\textbf{$\langle k \rangle$} & 
\textbf{$\langle L \rangle$} & 
\textbf{$CC$} & 
\textbf{$d$} & 
\textbf{$r$} \\
\midrule
Kenyan-Households-Across & 21 & 54 & 5.14 & 1.90 & 0.21 & 4 & -0.65 \\
Taro-exchange & 22 & 39 & 3.55 & 2.49 & 0.28 & 5 & -0.38 \\
Bison & 26 & 222 & 17.08 & 1.32 & 0.79 & 2 & -0.10 \\
Sheep & 28 & 235 & 16.79 & 1.38 & 0.73 & 3 & -0.00 \\
soc-firm-hi-tech & 33 & 91 & 5.52 & 2.36 & 0.37 & 5 & -0.09 \\
soc-karate & 34 & 78 & 4.59 & 2.41 & 0.26 & 5 & -0.48 \\
road-chesapeake & 39 & 170 & 8.72 & 1.84 & 0.28 & 3 & -0.38 \\
HIV & 40 & 41 & 2.05 & 4.47 & 0.03 & 10 & -0.28 \\
ENZYMES-g14 & 41 & 73 & 3.56 & 8.58 & 0.52 & 23 & 0.13 \\
contiguous-usa & 49 & 107 & 4.37 & 4.16 & 0.41 & 11 & 0.23 \\
Japanses-macaques & 62 & 1167 & 37.65 & 1.38 & 0.66 & 2 & -0.07 \\
soc-dolphins & 62 & 159 & 5.13 & 3.36 & 0.31 & 8 & -0.04 \\
Train-bombing & 64 & 243 & 7.59 & 2.69 & 0.56 & 6 & 0.03 \\
cat-mixed-species-brain-1 & 65 & 730 & 22.46 & 1.70 & 0.57 & 3 & -0.03 \\
edit-gotwikibooks & 65 & 92 & 2.83 & 3.87 & 0.04 & 9 & -0.42 \\
Highschool & 70 & 274 & 7.83 & 2.64 & 0.40 & 6 & 0.08 \\
ca-sandi-auths & 86 & 124 & 2.88 & 4.81 & 0.27 & 11 & -0.26 \\
ENZYMES-g297 & 121 & 149 & 2.46 & 10.96 & 0.10 & 29 & 0.16 \\
ENZYMES-g295 & 123 & 139 & 2.26 & 12.21 & 0.03 & 31 & 0.21 \\
ENZYMES-g296 & 125 & 141 & 2.26 & 12.94 & 0.03 & 32 & 0.29 \\
edit-dinwiki & 171 & 553 & 6.47 & 2.17 & 0.09 & 4 & -0.59 \\
convote & 219 & 521 & 4.76 & 3.31 & 0.12 & 7 & -0.34 \\
Physicians & 241 & 923 & 7.66 & 2.59 & 0.25 & 5 & -0.06 \\
econ-wm3 & 257 & 2379 & 18.51 & 2.61 & 0.58 & 6 & -0.06 \\
econ-wm1 & 258 & 2389 & 18.52 & 2.94 & 0.57 & 11 & -0.12 \\
econ-wm2 & 259 & 2533 & 19.56 & 2.62 & 0.58 & 6 & 0.02 \\
inf-USAir97 & 332 & 2126 & 12.81 & 2.74 & 0.40 & 6 & -0.21 \\
ca-netscience & 379 & 914 & 4.82 & 6.04 & 0.43 & 17 & -0.08 \\
bio-celegans-dir & 453 & 2040 & 9.01 & 2.66 & 0.12 & 7 & -0.22 \\
power-494-bus & 494 & 586 & 2.37 & 10.47 & 0.05 & 26 & -0.07 \\
fb-pages-food & 620 & 2102 & 6.78 & 5.09 & 0.22 & 17 & -0.03 \\
bio-DM-LC & 658 & 1129 & 3.43 & 6.61 & 0.06 & 18 & -0.12 \\
power-662-bus & 662 & 906 & 2.74 & 10.24 & 0.08 & 25 & -0.07 \\
power-685-bus & 685 & 1282 & 3.74 & 12.42 & 0.21 & 26 & 0.18 \\
soc-wiki-Vote & 889 & 2914 & 6.56 & 4.10 & 0.13 & 13 & -0.03 \\
power-1138-bus & 1138 & 1458 & 2.56 & 12.72 & 0.09 & 31 & -0.08 \\
inf-euroroad & 1174 & 1417 & 2.41 & 18.40 & 0.03 & 62 & 0.13 \\
road-euroroad & 1174 & 1417 & 2.41 & 18.40 & 0.03 & 62 & 0.13 \\
maayan-faa & 1226 & 2408 & 3.93 & 5.93 & 0.06 & 17 & -0.02 \\
econ-mahindas & 1258 & 7513 & 11.94 & 3.57 & 0.04 & 8 & -0.06 \\
bio-CE-LC & 1387 & 1648 & 2.38 & 7.92 & 0.04 & 22 & -0.17 \\
bio-yeast & 1458 & 1948 & 2.67 & 6.81 & 0.05 & 19 & -0.21 \\
netscience & 1461 & 2742 & 3.75 & 6.04 & 0.69 & 17 & 0.46 \\
power-bcspwr09 & 1723 & 2394 & 2.78 & 15.49 & 0.08 & 37 & -0.10 \\
ca-CSphd & 1882 & 1740 & 1.85 & 11.75 & 0.00 & 28 & -0.20 \\
\bottomrule
\end{tabular}
\end{table}

\begin{figure}[h]
    \centering
	\centerline{\includegraphics[width=0.6\linewidth]{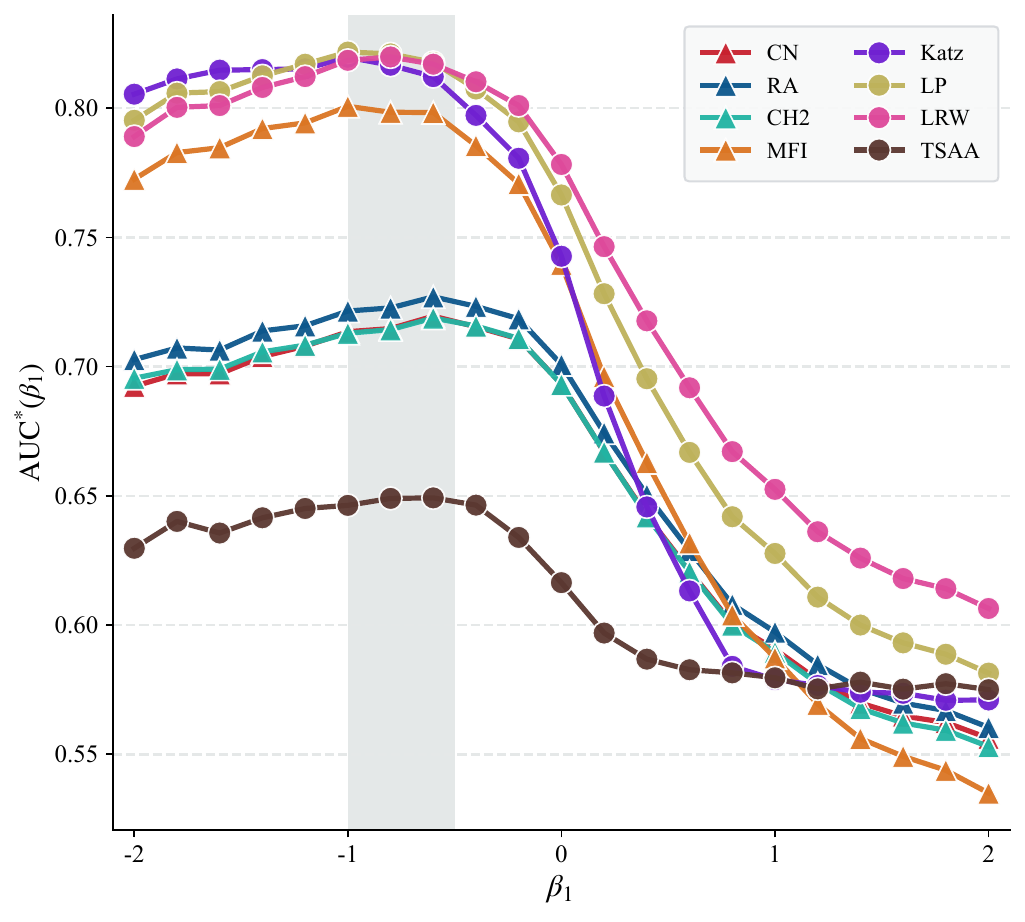}}
	\caption{
		\textbf{Best achievable AUC under different $\beta_1$.} Each curve reports $AUC^*(\beta_1)$ for one algorithm. All results are averaged over $45$ real-world networks, with $10$ independent runs for each network.}
	\label{fig_beta1_maxAUC}
\end{figure}

\section{Results}
To assess whether the structural characteristics of links in the probe set affect link prediction performance, we vary $\beta_1$ in the first-stage sampling to generate different probe sets that can be considered as proxies of true missing links. Given an algorithm, let $AUC(\beta_1,\beta_2)$ denote the AUC value when the two sampling parameters are $\beta_1$ and $\beta_2$, respectively. For a fixed $\beta_1$, we define
\begin{equation}
AUC^{*}(\beta_1)=\max_{\beta_2} AUC(\beta_1,\beta_2),
\end{equation}
which represents the highest AUC achieved across all $\beta$-sampling strategies when the structural distribution of the probe set is statistically determined by the parameter $\beta_1$. As shown in \autoref{fig_beta1_maxAUC}, when $\beta_1$ varies, $AUC^{*}(\beta_1)$ changes markedly for all considered algorithms, indicating that the best achievable performance strongly depends on the structural distribution of the probe set, and thus we can confidently infer that missing patterns exert a substantial impact on the performance of link prediction algorithms.

An additional notable observation is that the $AUC^{*}(\beta_1)$ curves of different algorithms generally exhibit a unimodal shape, with peaks concentrated in the interval $[-1.0,-0.5]$, suggesting that predictive performance tends to be higher when links in the probe set are likely to be associated with high-degree nodes. If $\beta_1>0$, links associated with high-degree nodes are more likely to be retained in $E^T \cup E^V$, and thus the probe set becomes biased toward links associated with low-degree nodes, which usually provide less neighborhood evidence and are therefore harder to be recovered, resembling the well-known difficult in the cold-start problem in recommender systems \cite{schein2002, lam2008, zhang2010, lika2014}. Accordingly, it is not surprising that the peak of $AUC^{*}(\beta_1)$ is achieved when $\beta_1<0$. At the same time, if $\beta_1$ is too small, very few links connecting two high-degree nodes will belong to $E^T \cup E^V$, leaving insufficient information for model training and thus degrading the predictive capacity of the model. Such trade-off leads to the observed peaks within the interval $[-1.0,-0.5]$.

\begin{figure}[htbp]
    \centering
	\centerline{\includegraphics[width=1.0\linewidth]{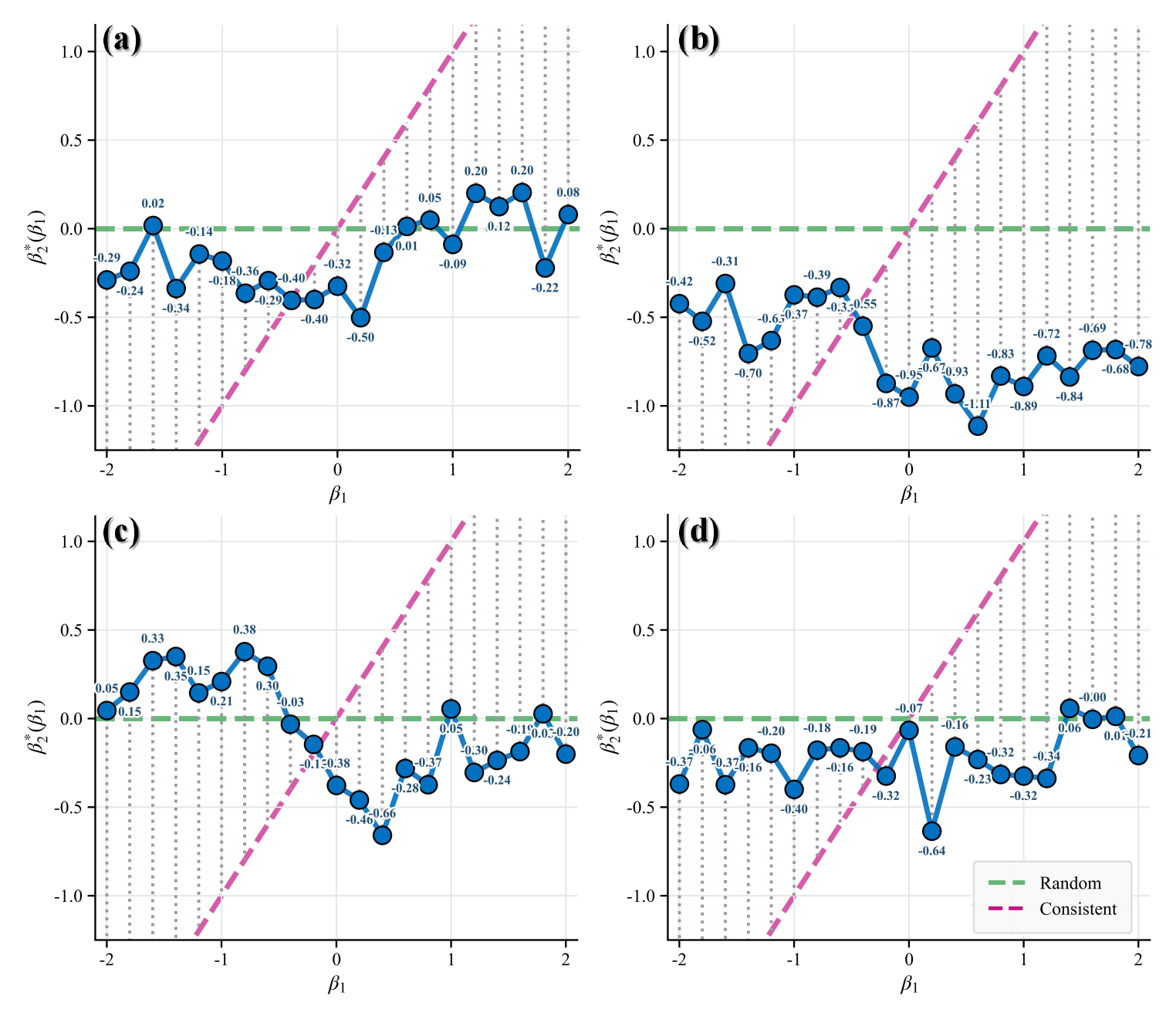}}
	\caption{
	\textbf{Values of $\beta_2^{*}(\beta_1)$ for parameterized algorithms.} The blue solid line with circles and marked values represent $\beta_2^{*}(\beta_1)$, while the green and magenta dash lines correspond to the random sampling $(\beta_2=0)$ and consistent sampling $(\beta_2=\beta_1)$. (a)-(d) report the results for Katz, LP, LRW, and TSAA, respectively.}
	\label{fig_beta1_optimalBeta2}
\end{figure}

\newcommand{\completenessbar}[5]{
    \begin{tikzpicture}[baseline=0ex]
        \fill[#3!15, rounded corners=1pt] (0, 0) rectangle (#2, 1.2ex);
        \fill[#3, rounded corners=1pt] (0, 0) rectangle (#1, 1.2ex);
        \node[anchor=south, inner sep=0pt, font=\small] at (#2/2, 1.6ex) {#4};
        \node[anchor=west, inner sep=0pt, font=\footnotesize, text=black!85] at (#2, 0.6ex) {#5};
    \end{tikzpicture}
}

\begin{table}[htbp]
\centering
\caption{
    \textbf{Performance comparison for the three sampling strategies.}
    Five values of $\beta_1$ (-2.0,-1.0,0,1.0,2) are considered, and for each fixed $\beta_1$, performance measured by AUC is reported under random sampling $(\beta_2=0)$, consistent sampling $(\beta_2=\beta_1)$, and optimal sampling $(\beta_2=\beta_2^*(\beta_1))$. The horizontal bars provide a normalized visual representation of the AUC on a common scale, with darker segments denoting the achieved performance and lighter segments indicating the remaining portion of the display scale. The two percentages in parentheses in the final column quantify the relative improvements of optimal sampling over random sampling and consistent sampling, respectively computed as
    $(AUC_{\mathrm{opt}}-AUC_{\mathrm{rand}})/AUC_{\mathrm{rand}}\times100\%$ and
    $(AUC_{\mathrm{opt}}-AUC_{\mathrm{cons}})/AUC_{\mathrm{cons}}\times100\%$, where $AUC_{\mathrm{opt}}=\max_{\beta_2}AUC(\beta_1,\beta_2)$,
    $AUC_{\mathrm{rand}}=AUC(\beta_1,0)$, and
    $AUC_{\mathrm{cons}}=AUC(\beta_1,\beta_1)$.
}
\label{tab:advanced_auc}
\renewcommand{\arraystretch}{2.2}
\resizebox{\textwidth}{!}{%
\begin{tabular}{l | c | l c l c l}
\toprule
\textbf{Algorithm} & \textbf{$\beta_1$} &
\textbf{1. Random ($\beta_2=0$)} & &
\textbf{2. Consistent ($\beta_2=\beta_1$)} & &
\textbf{3. Optimal ($\beta_2=\beta_2^{*}(\beta_1)$)} \\
\midrule
\multirow{5}{*}{\textbf{Katz}} & -2.0 & \completenessbar{1.742}{2.5}{bar_random}{0.803}{} & & \completenessbar{0.808}{2.5}{bar_consist}{0.560}{} & & \completenessbar{1.752}{2.5}{bar_optimal}{\textbf{0.805}}{\hspace{0.2cm} (+0.3\% $\nearrow$, +43.8\% $\nearrow$)} \\
 & -1.0 & \completenessbar{1.791}{2.5}{bar_random}{0.816}{} & & \completenessbar{1.769}{2.5}{bar_consist}{0.810}{} & & \completenessbar{1.807}{2.5}{bar_optimal}{\textbf{0.820}}{\hspace{0.2cm} (+0.5\% $\nearrow$, +1.2\% $\nearrow$)} \\
 & 0.0 & \completenessbar{1.503}{2.5}{bar_random}{0.741}{} & & \completenessbar{1.503}{2.5}{bar_consist}{0.741}{} & & \completenessbar{1.510}{2.5}{bar_optimal}{\textbf{0.743}}{\hspace{0.2cm} (+0.2\% $\nearrow$, +0.2\% $\nearrow$)} \\
 & +1.0 & \completenessbar{0.830}{2.5}{bar_random}{0.566}{} & & \completenessbar{0.874}{2.5}{bar_consist}{0.577}{} & & \completenessbar{0.880}{2.5}{bar_optimal}{\textbf{0.579}}{\hspace{0.2cm} (+2.3\% $\nearrow$, +0.3\% $\nearrow$)} \\
 & +2.0 & \completenessbar{0.684}{2.5}{bar_random}{0.528}{} & & \completenessbar{0.680}{2.5}{bar_consist}{0.527}{} & & \completenessbar{0.850}{2.5}{bar_optimal}{\textbf{0.571}}{\hspace{0.2cm} (+8.2\% $\nearrow$, +8.4\% $\nearrow$)} \\
\midrule
\multirow{5}{*}{\textbf{TSAA}} & -2.0 & \completenessbar{0.687}{2.5}{bar_random}{0.529}{} & & \completenessbar{1.053}{2.5}{bar_consist}{0.624}{} & & \completenessbar{1.076}{2.5}{bar_optimal}{\textbf{0.630}}{\hspace{0.2cm} (+19.1\% $\nearrow$, +1.0\% $\nearrow$)} \\
 & -1.0 & \completenessbar{0.723}{2.5}{bar_random}{0.538}{} & & \completenessbar{0.809}{2.5}{bar_consist}{0.560}{} & & \completenessbar{1.140}{2.5}{bar_optimal}{\textbf{0.646}}{\hspace{0.2cm} (+20.1\% $\nearrow$, +15.3\% $\nearrow$)} \\
 & 0.0 & \completenessbar{0.832}{2.5}{bar_random}{0.566}{} & & \completenessbar{0.832}{2.5}{bar_consist}{0.566}{} & & \completenessbar{1.025}{2.5}{bar_optimal}{\textbf{0.616}}{\hspace{0.2cm} (+8.9\% $\nearrow$, +8.9\% $\nearrow$)} \\
 & +1.0 & \completenessbar{0.869}{2.5}{bar_random}{0.576}{} & & \completenessbar{0.846}{2.5}{bar_consist}{0.570}{} & & \completenessbar{0.883}{2.5}{bar_optimal}{\textbf{0.580}}{\hspace{0.2cm} (+0.6\% $\nearrow$, +1.7\% $\nearrow$)} \\
 & +2.0 & \completenessbar{0.862}{2.5}{bar_random}{0.574}{} & & \completenessbar{0.865}{2.5}{bar_consist}{0.575}{} & & \completenessbar{0.865}{2.5}{bar_optimal}{\textbf{0.575}}{\hspace{0.2cm} (+0.1\% $\nearrow$, +0.0\% $\nearrow$)} \\
\midrule
\multirow{5}{*}{\textbf{LP}} & -2.0 & \completenessbar{1.693}{2.5}{bar_random}{0.790}{} & & \completenessbar{1.710}{2.5}{bar_consist}{0.795}{} & & \completenessbar{1.713}{2.5}{bar_optimal}{\textbf{0.795}}{\hspace{0.2cm} (+0.6\% $\nearrow$, +0.1\% $\nearrow$)} \\
 & -1.0 & \completenessbar{1.798}{2.5}{bar_random}{0.818}{} & & \completenessbar{1.805}{2.5}{bar_consist}{0.819}{} & & \completenessbar{1.814}{2.5}{bar_optimal}{\textbf{0.822}}{\hspace{0.2cm} (+0.5\% $\nearrow$, +0.3\% $\nearrow$)} \\
 & 0.0 & \completenessbar{1.593}{2.5}{bar_random}{0.764}{} & & \completenessbar{1.593}{2.5}{bar_consist}{0.764}{} & & \completenessbar{1.602}{2.5}{bar_optimal}{\textbf{0.766}}{\hspace{0.2cm} (+0.3\% $\nearrow$, +0.3\% $\nearrow$)} \\
 & +1.0 & \completenessbar{1.068}{2.5}{bar_random}{0.628}{} & & \completenessbar{1.045}{2.5}{bar_consist}{0.622}{} & & \completenessbar{1.068}{2.5}{bar_optimal}{\textbf{0.628}}{\hspace{0.2cm} (+0.0\% $\nearrow$, +1.0\% $\nearrow$)} \\
 & +2.0 & \completenessbar{0.890}{2.5}{bar_random}{0.581}{} & & \completenessbar{0.890}{2.5}{bar_consist}{0.581}{} & & \completenessbar{0.890}{2.5}{bar_optimal}{\textbf{0.581}}{\hspace{0.2cm} (0.0\%, 0.0\%)} \\
\midrule
\multirow{5}{*}{\textbf{LRW}} & -2.0 & \completenessbar{1.678}{2.5}{bar_random}{0.786}{} & & \completenessbar{1.687}{2.5}{bar_consist}{0.788}{} & & \completenessbar{1.688}{2.5}{bar_optimal}{\textbf{0.789}}{\hspace{0.2cm} (+0.3\% $\nearrow$, +0.1\% $\nearrow$)} \\
 & -1.0 & \completenessbar{1.792}{2.5}{bar_random}{0.816}{} & & \completenessbar{1.795}{2.5}{bar_consist}{0.817}{} & & \completenessbar{1.802}{2.5}{bar_optimal}{\textbf{0.819}}{\hspace{0.2cm} (+0.3\% $\nearrow$, +0.2\% $\nearrow$)} \\
 & 0.0 & \completenessbar{1.643}{2.5}{bar_random}{0.777}{} & & \completenessbar{1.643}{2.5}{bar_consist}{0.777}{} & & \completenessbar{1.647}{2.5}{bar_optimal}{\textbf{0.778}}{\hspace{0.2cm} (+0.1\% $\nearrow$, +0.1\% $\nearrow$)} \\
 & +1.0 & \completenessbar{1.162}{2.5}{bar_random}{0.652}{} & & \completenessbar{1.149}{2.5}{bar_consist}{0.649}{} & & \completenessbar{1.163}{2.5}{bar_optimal}{\textbf{0.652}}{\hspace{0.2cm} (+0.1\% $\nearrow$, +0.6\% $\nearrow$)} \\
 & +2.0 & \completenessbar{0.986}{2.5}{bar_random}{0.606}{} & & \completenessbar{0.986}{2.5}{bar_consist}{0.606}{} & & \completenessbar{0.986}{2.5}{bar_optimal}{\textbf{0.606}}{\hspace{0.2cm} (0.0\%, +0.0\% $\nearrow$)} \\
\bottomrule
\end{tabular}}
\end{table}

\begin{figure}[h]
    \centering
	\centerline{\includegraphics[width=0.9\linewidth]{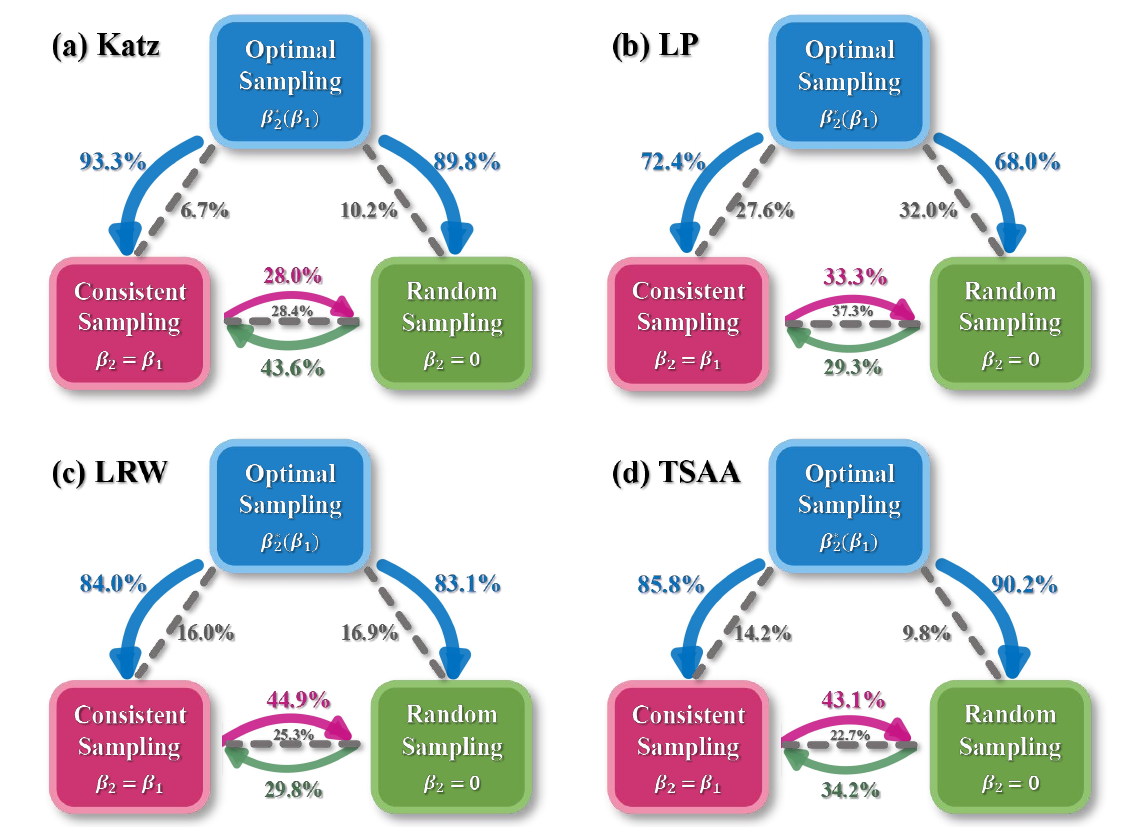}}
	\caption{
	\textbf{Pairwise comparison of prediction performance among the three sampling strategies.} Each node represents one sampling strategy, and the value associated with a directed edge from strategy \textbf{A} to strategy \textbf{B} indicates the winning rate that \textbf{A} outperforms \textbf{B} in total $225$ pairwise comparisons ($45$ networks and $5$ representative $\beta_1$ values). Gray dashed bidirectional edges denote the proportion of cases in which two strategies achieve equal AUC. (a)-(d) show the results for Katz, LP, LRW, and TSAA, respectively.
	}
    \label{fig_compare}
\end{figure}

Having shown that $\beta_1$ plays a critical role, we next investigate which strategy (\textit{i.e.}, which value of $\beta_2$) achieves the best performance given a fixed 
$\beta_1$, a specific prediction algorithm, and the $\beta$-sampling scheme. Given $\beta_1$, we define 
\begin{equation}
\beta_2^*(\beta_1)=\arg\max_{\beta_2} AUC(\beta_1,\beta_2).
\end{equation}
Since random sampling and consistent sampling correspond to $\beta_2=0$ and $\beta_2=\beta_1$, $\beta_2^{*}(\beta_1)$ should be close to zero if random sampling yields the best performance, whereas it should take a value similar to $\beta_1$ when consistent sampling performs best. As shown in \autoref{fig_beta1_optimalBeta2}, The trajectories of $\beta_2^{*}(\beta_1)$ deviate from both reference cases (\textit{i.e.}, $\beta_2=0$ and $\beta_2=\beta_1$), indicating that random sampling does not yield the best performance, and to make the validation set structurally similar to the probe set also does not achieve the best performance. Table \ref{tab:advanced_auc} compares the three
representative sampling strategies: random sampling $(\beta_2=0)$, consistent sampling $(\beta_2=\beta_1)$, and optimal sampling $(\beta_2=\beta_2^{*}(\beta_1))$. We first focus on how much the optimal strategy improves upon random sampling and consistent sampling. For LP and LRW, the gains are generally small, while for some cases of TSAA and Katz, the improvement is pronounced. This result suggests that the benefit of optimizing $\beta_2$ strongly depends on both $\beta_1$ and the adopted algorithm. We next concern whether consistent sampling outperforms random sampling. As shown in Table \ref{tab:advanced_auc}, the two sampling strategies are competitive. We further compare the performance of the three aforementioned sampling strategies across each network individually. For each network, we evaluate the five representative values of $\beta_1$ listed in Table \ref{tab:advanced_auc}. This results in a total of $45 \times 5=225$ pairwise comparisons, where each comparison relies on the average AUC values obtained from $10$ independent runs. As shown in Figure \ref{fig_compare}, consistent sampling has a slightly higher winning rate than random sampling for LRW and TSAA, and a slightly lower winning rate for Katz. While for LP, their winning rates are more or less the same. These results suggest that the two baselines are broadly comparable, with their relative advantage depending on the algorithm rather than following a universal rule. To briefly summarize, our results demonstrate that the second-stage sampling still affects the final prediction performance even with a fixed $\beta_1$. Moreover, the optimal choice of $\beta_2$ is non-trivial. In other words, neither the widely adopted random sampling nor intuitively promising consistent sampling yields optimal performance. Surprisingly, consistent sampling barely outperforms random sampling; any marginal improvement it brings is extremely limited, if not negligible. These findings indicate that we should not regard the second-stage sampling merely as an auxiliary operation. Instead, we need to explore how to optimize this sampling procedure to achieve better predictions. Such optimization requires consideration of two aspects: the similarity between the validation set and the probe set, and whether the edges retained in the training set contain sufficient informative signals to facilitate model training.

\section{Conclusion and Discussion}
In network science, sampling has long been studied, with most existing works focusing on enhancing visualization, preserving network statistics, retaining specific structural patterns, and improving computational efficiency \cite{rafiei2005, stumpf2005, leskovec2006, gjoka2010, maiya2010, de2010}. In comparison, much less attention has been paid to how sampling strategies affect the difficulty and thus performance of graph mining tasks, as well as whether and how we can improve algorithmic performance by designing sampling strategies (see \cite{lichtnwalter2012, mitchell2021} for some relevant but not directly related discussions). 

This paper focuses on the link prediction task and systematically investigates how link sampling strategies affect algorithmic performance. By proposing a sampling strategy with continuously tunable structural features for sampled links and treating the partition of link set $E$ as two sampling processes, we implement extensive experiments and draw two important conclusions: (i) The structural characteristics of missing links, as simulated by varying probe sets, substantially affect prediction accuracy, with missing links that tend to connect high-degree nodes easier to predict; (ii) The second-stage sampling has a significant influence on prediction accuracy, and the optimal strategy under the $\beta$-sampling framework is neither random sampling nor consistent sampling.

\begin{figure}[h]
    \centering
	\centerline{\includegraphics[width=0.65\linewidth]{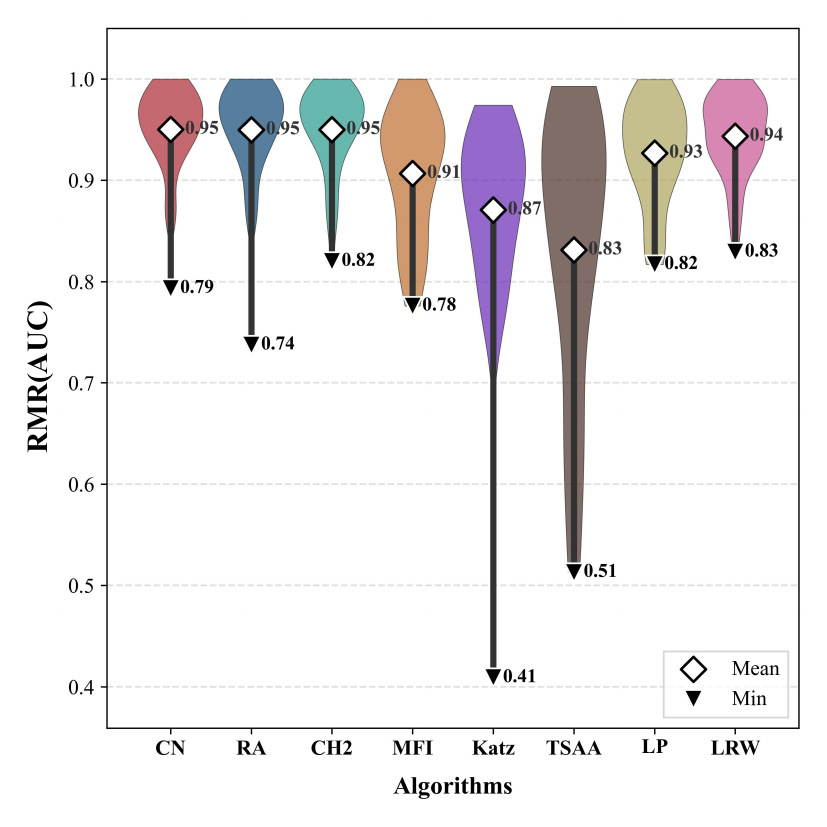}}
	\caption{
		\textbf{Random-to-maximal Ratios (RMRs) of the eight link prediction algorithms based on AUC.} Different colored boxes on the $x$-axis correspond to different link prediction algorithms, and the $y$-axis reports RMR values. For each algorithm, the distribution is obtained over the $45$ real-world networks. The hollow diamond inside each box denotes the mean RMR, and the triangle denotes the minimum value.
	}
	\label{fig_ceilingscore_AUC}
\end{figure}

The above findings show that link prediction performance is shaped not only by the algorithm, but also by the missing pattern and how the remaining links are divided (\textit{i.e.}, which links belong to $E^T$, and which belong to $E^V$). However, most known studies only consider the trivial case corresponding to $\beta_2=\beta_1=0$. When we claim that algorithm \textbf{X} achieves higher prediction accuracy than algorithm \textbf{Y}, we merely mean that \textbf{X} obtains a higher $AUC(0,0)$ than \textbf{Y} (the same logic applies to other evaluation metrics). However, for other values of $\beta_1$ and $\beta_2$, \textbf{X} may yields inferior performance compared with \textbf{Y}. How large is the gap between the $AUC(0,0)$ adopted in conventional studies and the maximum AUC achievable under different combinations of $\beta_1$ and $\beta_2$? To address this issue, we propose a measure called \textbf{Random-to-maximal Ratio} (RMR), defined as:
\begin{equation}
RMR=\frac{AUC(0,0)}{AUC^{*}},
\end{equation}
where 
\begin{equation}
AUC^{*}=\max_{\beta_1,\beta_2}AUC(\beta_1,\beta_2)
\end{equation}
is the maximum AUC value. If the RMR value of an algorithm is significantly less than $1$, this indicates that, under specific missing patterns and elaborate sampling strategy designs, the prediction performance of the algorithm can substantially exceed its performance under the default setting (\textit{i.e.}, $\beta_2=\beta_1=0$). As shown in \autoref{fig_ceilingscore_AUC}, the average $\mathrm{RMR}$ is usually between $0.80$ and $0.95$, but in some cases, the RMR value can be even lower than $0.5$, indicating that relying solely on $AUC(0,0)$ to evaluate algorithmic performance is far from sufficient, therefore we have to gain an in-depth and comprehensive understanding of the entire link sampling procedure and treat it as a critical component for algorithm optimization.

Two limitations should be noted. First, $\beta$-sampling is based on the degrees of the two endpoints, while real missing links may have nontrivial structural features related to motif and community membership, role positions, path dependence, spatial constraints, temporal recency, and higher-order interaction patterns. Future work could extend $\beta$-sampling or develop alternative tunable sampling schemes that account for these additional features. Second, the present analysis is only based on simple networks, and the future work should investigate more complex network settings, such as weighted networks \cite{murata2007, lu2010, zhao2015}, temporal networks \cite{dunlavy2011, holme2012, tang2020}, and higher-order networks \cite{pujari2015, benson2018, bian2025}. 

In summary, evaluation in link prediction should move beyond the habitual use of default random sampling and pay closer attention to the structural patterns of missing links, the effect of sampling bias, and the performance variation induced by different sampling strategies.

\section*{Author contributions}
Y.B. and T.Z. designed the study. Y.B. performed the research. Y.B. prepared the figures. Z.D. checked and improved the code. X.J. provided the code for several link prediction algorithms. Y.B., Z.D., X.J. and T.Z. analyzed the data. Y.B. and T.Z. wrote the manuscript.

\section*{Competing interests}
The authors declare no competing interests.

\bibliographystyle{plain}

\end{document}